\documentclass[runningheads]{llncs}
\usepackage{graphicx}
\usepackage{amsmath}
\usepackage{amssymb}
\usepackage{booktabs}
\usepackage{graphicx}
\usepackage{comment}
\usepackage{mdwlist}
\usepackage{paralist}
\usepackage{colortbl}
\usepackage{tabularx}
\usepackage[colorinlistoftodos]{todonotes}
\usepackage{hyperref}
\usepackage{float}
\usepackage{algorithm2e}
\usepackage{subcaption}
\usepackage[export]{adjustbox}
\usepackage{xcolor} 
\usepackage{enumitem}
\usepackage{todonotes}
\usepackage{multirow}

\newcommand{\nores}{\cellcolor{lightgray}}

\newcommand{\lmjm}{LMJM}
\newcommand{\lmdr}{LMDir}
\newcommand{\para}[1]{\textbf{#1}.}

\newcommand{\pr}{Pearson's-$\rho$}

\definecolor{magicmint}{rgb}{0.67, 0.94, 0.82}
\definecolor{melon}{rgb}{0.99, 0.74, 0.71}

\begin{document}

\title{An Analysis of Variations in the Effectiveness of Query Performance Prediction}%
\titlerunning{An Analysis of Variations in the Effectiveness of QPP}

\author{
Debasis Ganguly
\inst{1}
\orcidID{0000-0003-0050-7138} \and
Suchana Datta
\inst{2}
\orcidID{0000-0001-9220-6652} \and
Mandar Mitra
\inst{3}
\orcidID{0000-0001-9045-9971} \and
Derek Greene
\inst{2}
\orcidID{0000-0001-8065-5418}
}
\authorrunning{D. Ganguly et al.}
%
\institute{University of Glasgow, United Kingdom \and
University College Dublin, Ireland \and
Indian Statistical Institute, Kolkata, India\\
\email{debasis.ganguly@glasgow.ac.uk},
\email{suchana.datta@ucdconnect.ie},
\email{mandar@isical.ac.in},
\email{derek.greene@ucd.ie}}

\maketitle              
\begin{abstract}
\label{abstract}
A query performance predictor estimates the retrieval effectiveness of a
system for a given query. Query performance prediction (QPP) algorithms are themselves
evaluated by measuring the correlation between the predicted effectiveness
and the actual effectiveness of a system for a set of queries. This
generally accepted framework for judging the usefulness of a QPP method
includes a number of sources of variability. For example, ``actual
effectiveness'' can be measured
using different metrics, 
for different rank cut-offs. 
The objective of this study is to identify some of these sources, and
investigate how variations in the framework can affect the outcomes of QPP
experiments. We consider this issue not only in terms of the absolute values of
the evaluation metrics being reported (e.g., Pearson's $r$, Kendall's
$\tau$), but also with respect to the changes in the ranks of different QPP
systems when ordered by the QPP metric scores.
Our experiments reveal that the observed QPP outcomes can vary considerably, both in terms of the absolute evaluation metric values and also in terms of the relative system ranks. We report the combinations of QPP evaluation metric and experimental settings that are likely to lead to smaller variations in the observed results\footnote{To appear in the proceedings of $44$th European Conference on Information Retrieval (ECIR'22).}.

\end{abstract}

\section{Introduction}
\label{sec:intro}
The problem of \emph{query performance prediction} (QPP)
\cite{carmel_sigir10,croft_qpp_sigir02,precision_prediction,survey_preret_qpp,postret_ictir11,uef_kurland_sigir10,qpp_ref_list_sigir17,kurland_tois12,qpp_croft_cikm06,wig_croft_SIGIR07}
has attracted the attention of the Information Retrieval (IR) community
over a number of years. QPP involves estimating the retrieval quality of an
IR system. 
A diverse range of pre-retrieval
(e.g. avgIDF \cite{survey_preret_qpp}) and post-retrieval approaches (e.g.
WIG \cite{wig_croft_SIGIR07}, NQC \cite{kurland_tois12}, UEF
\cite{uef_kurland_sigir10}) have been proposed for the task of QPP. 

The primary use case of QPP can be described as follows: ``If we could
determine in advance which retrieval approach would work well for a given
query, then hopefully, selecting the appropriate retrieval method on a
[per] query basis could improve the retrieval effectiveness
significantly.''~\cite{carmel2010monograph}. In other words, the objective
of QPP would be to predict how easy or difficult a given query is for an IR
system. This prediction could either be a categorical label (e.g.,
\textsc{easy}, \textsc{moderate}, \textsc{hard}), or a numerical estimate
of a standard IR evaluation metric (which generally lies in $[0,1]$).

QPP is a challenging problem, however, and this eventual objective has
remained elusive thus far. Given a query and an IR system, well-known QPP
methods simply compute a real-valued score that is meant to be indicative
of the effectiveness of the system for the given query.
While this score is typically not interpreted as a statistical estimate of
a specific evaluation metric (e.g. AP or nDCG~\cite{ndcg}), it is expected
to be highly correlated with a standard evaluation measure. Indeed, the
quality of a QPP method is usually determined by measuring the correlation
between its predicted effectiveness scores and the values of some standard
evaluation metric for a set of queries.

Consider a proposed QPP algorithm $\mathcal{P}$. Given an IR system $S$,
and a set of queries $\mathcal{Q} = \{Q_1, Q_2, \ldots, Q_n\}$, $S$ is used
to retrieve a ranked list $L_i$ of documents for each $Q_i \in
\mathcal{Q}$. For each $L_i$, $\mathcal{P}$ computes a predicted
effectiveness score $\phi_i$. Using available relevance assessments as
ground-truth, a standard IR metric $g_i$ is also computed for $L_i$. The
correlation between the lists $\{\phi_1, \phi_2, \ldots, \phi_n\}$ and
$\{g_1, g_2, \ldots, g_n\}$ is taken to be a measure of how effective
$\mathcal{P}$ is as a query performance predictor.

In this study, we analyse the above approach for evaluating and comparing
different QPP methods. We identify the sources of variability within this
generally accepted framework, and show that these variations can lead to
differences in the computed correlations. This, in turn, can lead to
differences in
\begin{compactitem}
\item the absolute values of reported QPP evaluation measures (e.g., the
  $\rho$ value for NQC \cite{kurland_tois12} measured with AP@100 as the
  target metric and LM-Dirichlet as the retrieval model can be
  substantially different from that measured with AP@1000 as the target
  metric and BM25 as the retrieval model on the \emph{same} set of
  queries); and also in 
\item the comparative effectiveness of a number of different QPP measures
  (e.g., NQC turns out to be better than WIG with AP@100, whereas WIG
  outperforms NQC when QPP effectiveness is measured using nDCG@10).
\end{compactitem}
Thus, these variations can lead to difficulties in reproducing QPP results,
both at the level of the correlation values being reported, and also in
terms of the relative performance of different competing methods on
standard datasets.


\para{Contributions} We conduct a range of experiments to analyze the
potential variations in QPP effectiveness results under different
experimental conditions. Specifically, we consider different combinations
of IR metrics and IR models (as well as rank cut-off values). The
experiments described in Section \ref{sec:res} reveal that the results of
QPP depend significantly on these settings. Thus, it may be difficult to
reproduce QPP experiments without a precise description of the experimental
context. While variations in other factors, such as the choice of indexing
implementation and set of pre-processing steps, may also matter, we
recommend that any empirical study of QPP include a precise description of
at least the above experimental settings in order to reduce variations in
reported results. More importantly, our findings suggest that it may even be
worthwhile to systematically revisit reported comparisons between competing QPP
approaches.




\section{Related Work}
\label{sec:rel}
Analyzing the sensitivity of reported results on the experiment settings is important for an empirical discipline such as IR. Buckley and Voorhees while examining the stability of commonly used evaluation measures in IR \cite{buckley_eval_stability}, reported observations, such as $\mathrm{P}@30$ has about twice the average error rate as compared to average precision (AP), or that a stable measurement of $\mathrm{P}@10$ requires an aggregation of over $50$ queries etc. 

Previous studies have investigated the sensitivity of relative ranks of IR systems to the pooling depth used for relevance assessments. It is reported that smaller samples of the relevance ground-truth obtained with smaller pool depths usually do not lead to significant changes in the relative performance of IR systems \cite{emine_inferred_ap,emine_estimate_ap,emine_estimate_ap_cikm06,buckley_incomplete_info}.
In relation to pooling, Buckley et. al. demonstrated that pools created during the TREC 2005 workshop exhibit a specific bias in favor of relevant documents, specifically contain the title words.

The study in \cite{dr_we_variation} analyzed the sensitivity of variations in embeddding vectors used for IR models.
The work in \cite{zobel_cikm09} stressed the importance of reproducibility in IR research by noting that most of the improvements reported over the years were not statistically significant over their predecessors.
Recently, this observation has also been reinforced for neural models by arguing that most of the neural approaches have compared their results against relatively weak baselines \cite{neural_hype_sigforum,neural_hype_sigir19}.


Somewhat similar to our investigation of the stability of QPP results relative to IR models and evaluation metrics, an inconsistency in QPP evaluation with respect to IR models and variations in query formulation was shown in \cite{scholer09,scholer17}.

\section{Anatomy of a QPP Evaluation Framework} 
\label{sec:rqs}
In this section, we formally define the various components in a standard QPP evaluation framework. 
As we demonstrate later, variations in these components can potentially lead to different experimental outcomes.

\begin{definition}
  \label{def:context}
  The context, $\mathcal{C}(Q)$, of a QPP experiment on a query $Q$, is a
  $3$-tuple of the form of $(\theta, \mathcal{S}, \kappa)$, where 
  $\kappa$ is a positive integer;
  the function $\mathcal{S}: Q\times D \mapsto \mathbb{R}$ is a scoring
  function that computes query-document similarities, and is used to
  retrieve $L = (D_1,\ldots,D_\kappa)$, the list of $\kappa$ top-ranked
  documents for $Q$ from a collection; and
  $\theta: L \mapsto [0,1]$ is an evaluation metric function that, given a
  query $Q$, a list $L$ of top-ranked documents, and $R(Q)$, the relevance
  assessments for $Q$, outputs a measure of usefulness of $L$.
\end{definition}

\begin{definition} \label{def:gt}
The ground-truth or reference value of retrieval effectiveness of a query $Q$ in relation to a QPP context, $\mathcal{C}(Q)$ of Definition \ref{def:context}, is a function of the form $g: \mathcal{C}(Q) \mapsto [0, 1]$. 
\end{definition}

\begin{definition}
  \label{def:qppmethod}
  A QPP method is a function of the form $\phi(Q, D_1,\ldots,D_k) \mapsto
  [0,1]$, which, given a query $Q$ and a list of top-$k$ retrieved
  documents\footnote{For pre-retrieval QPP approaches, $(D_1,\ldots,D_k) =
    \emptyset$.}, outputs a 
  number that is indicative of how relevant the retrieved list is. In other
  words, the output of the predictor $\phi(Q)$ is some measure of the
  ground-truth retrieval effectiveness measure $g(\mathcal{C}(Q))$ from
  Definition \ref{def:gt}.
\end{definition}
For example, NQC \cite{kurland_tois12} or WIG \cite{wig_croft_SIGIR07}
compute $\phi(Q)$ based on a set of $k$ top-ranked
documents\footnote{$k$ is a parameter of a post-retrieval QPP method, and
  can be different from $\kappa$, the number of top documents used for QPP
  evaluation.} and estimating how distinct it is from the rest of the
collection. The intuition behind NQC and WIG is that the higher the
distinctiveness, the higher the likelihood of finding more relevant
documents in the retrieved list.

The next step in QPP evaluation is to measure the correlation between the predicted retrieval effectiveness, $\phi(Q)$, and the ground-truth retrieval effectiveness, $g(\mathcal{C}(Q))$ over a set of benchmark queries $\mathcal{Q}$, 
using a correlation function, $\chi: (\Phi, \mathcal{G}(\mathcal{C})) \mapsto [0, 1]$, where 
$\Phi = \bigcup_{Q \in \mathcal{Q}} \phi(Q)$
and
$\mathcal{G}(\mathcal{C}) = \bigcup_{Q \in \mathcal{Q}}g(\mathcal{C}(Q)))$.
Common choices for $\chi$ are Pearson's $r$, which computes a correlation between the values themselves, and rank correlation measures, such as Spearman's $\rho$, which compute the correlation between the ordinals of the members of $\Phi$ and $\mathcal{G}(\mathcal{C})$. 


It is clear from Definitions \ref{def:context}-\ref{def:qppmethod} that the QPP outcome, $\chi(\Phi, \mathcal{G})(\mathcal{C})$, depends on the context $\mathcal{C}(Q)$ used for each $Q \in \mathcal{Q}$. Our first objective is to quantify the relative changes in QPP outcomes $\chi$ with changes in the context $\mathcal{C}(Q)$. In other words, we wish to compute the relative changes of the form
$|\chi(\Phi, \mathcal{G}(\mathcal{C}_i)) - \chi(\Phi, \mathcal{G}(\mathcal{C}_j))|$,
for two different instances of QPP contexts $\mathcal{C}_i = (\theta_i, \mathcal{S}_i, \kappa_i)$ and $\mathcal{C}_j = (\theta_j, \mathcal{S}_j, \kappa_j)$.
Thus, our first research question is the following:
\begin{itemize}
    \item[]\textbf{RQ1:} Do \textbf{variations in the QPP context}, $\mathcal{C}$, in terms of the IR metric ($\theta$), the IR model ($\mathcal{S}$) and the rank cut-off ($\kappa$) used to construct the QPP evaluation ground-truth, $g(\mathcal{C})$, lead to \textbf{significant differences in outcome of a QPP method} $\phi$?
\end{itemize}


\noindent 
Next, instead of computing the relative change in the outcome values (correlations) of individual QPP methods, we seek to measure the relative change in the rankings (in terms of effectiveness) of a number of different QPP methods. Formally, given a set of $m$ QPP functions $\{\phi_1,\ldots,\phi_m\}$, we compute the effectiveness of each with respect to a number of different QPP contexts, $\chi(\Phi_i,\mathcal{G}(\mathcal{C}_j))$ for $j=1,\ldots,n$. The objective is to investigate whether or not the ranking of QPP systems computed with different contexts is relatively stable. For instance, if NQC is the best method for a context that used LM-Dirichlet as retrieval model and AP@100 as evaluation metric, we might wish to investigate whether it remains the best method for a different QPP context, say, BM25 as the retrieval model and nDCG@10 as the evaluation metric.
Stated explicitly,
\begin{itemize}
    \item[]\textbf{RQ2:} Do \textbf{variations in the QPP context}, $\mathcal{C}$, in terms of the IR metric ($\theta$), the IR model ($\mathcal{S}$) and the rank cut-off ($\kappa$) used to construct the QPP evaluation ground-truth, $g(\mathcal{C})$, lead to \textbf{significant differences in the relative ranks of different QPP methods} $\phi_1,\ldots,\phi_m$?
\end{itemize}





\section{Experimental Setup}
\label{sec:exp}

To investigate the research questions from the last section, we conduct QPP experiments \footnote{Implementation available at: https://github.com/suchanadatta/qpp-eval.git} on a widely-used dataset, the TREC Robust dataset, which consists of 249 queries.
To address RQ1 and RQ2, we first define the set of possible QPP contexts that we explore in our experiments.

\begin{table}[t]
\small
\centering
\begin{tabular}{l@{~~~}l@{~~~}l@{~~~}c@{~~}c@{~~}c}
\toprule
Collection & \#Docs & Topic Set& \#Queries & Avg.$|Q|$ & Avg.\#Rel \\
\midrule
Disks 4,5 (w/o CR) & 528,155 & TREC-Robust & 249 &2.68 & 71.21 \\
\bottomrule
\end{tabular}
\caption{
\small
Characteristics of the TREC-Robust dataset used in our QPP experiments.
`Avg.$|Q|$' and `Avg.\#Rel' denote the average number of terms in a query, and the average number of relevant documents for a query, respectively.}
\label{tab:datastats}
\end{table}

\para{IR evaluation metrics investigated} As choices for the IR evaluation metric (i.e., the function $\theta$), we consider `AP', `nDCG', `P@10', and `recall'. The evaluation functions explored represent a mixture of both precision- and recall-oriented metrics. While AP and nDCG address both the aspects of precision and recall (leaning towards favouring precision), P@10 is a solely precision-oriented metric. To investigate RQ1, we set the cut-off for AP, nDCG, and recall to $100$, as is common in the literature on QPP \cite{kurland_tois12,hamed_neuralqpp,query_variants_kurland}.

\para{IR models investigated}
IR models represent the second component of a QPP context as per Definition \ref{def:context}. We explore three such models: a) language modeling with Jelinek-Mercer smoothing (\lmjm) \cite{lmdir,hiemstra}, b) language modeling with Dirichlet smoothing (\lmdr) \cite{lmdir}, and c) Okapi BM25 \cite{Okapi}.
The values of the IR model parameters were chosen after a grid search to optimize the MAP values on the TREC-Robust queries.
Unless otherwise specified, for \lmjm, we used $\lambda=0.6$, the value of $k_1$ and $b$ in BM25 were set to $0.7$ and $0.3$, respectively, and the value of the smoothing parameter $\mu$ for \lmdr~was set to $1000$.

\para{QPP methods tested}
To compare the relative perturbations in preferential ordering of the QPP systems in terms of the evaluated effectiveness, we employ a total of seven different QPP methods, as outlined below:
\begin{compactitem}
\item \textbf{AvgIDF} \cite{survey_preret_qpp} is a pre-retrieval QPP method that uses the average idfs of the constituent query terms as the predicted query performance estimate.

\item 
\textbf{Clarity} \cite{croft_qpp_sigir02} estimates a relevance model (RLM) \cite{Lavrenko_RLM2001:RBL:383952.383972} distribution of term weights from a set of top-ranked documents, and then computes its KL divergence with the collection model.

\item \textbf{WIG} \cite{wig_croft_SIGIR07} uses the aggregated value of the information gain of each document 
in the top-retrieved set as a specificity estimate.

\item \textbf{NQC} \cite{kurland_tois12} or
normalized query commitment estimates the specificity of a query as the standard deviation of the RSVs of the top-retrieved documents.

\item \textbf{UEF} \cite{uef_kurland_sigir10}
assumes that information from some top-retrieved sets of documents are more reliable than others.
A high perturbation of a ranked list after feedback indicates a poor retrieval effectiveness of the initial list. This, in turn, suggests that a smaller confidence should be associated with the QPP estimate of such a query.
Formally,
\begin{equation}
\mathrm{UEF}(Q,\phi) = \xi(R_M(Q), R_M(\theta_Q)) \phi(Q)
\end{equation}
where $\phi(Q)$ is the predicted score of a base QPP estimator (e.g. WIG or NQC), $R_M(\theta_Q)$ denotes the re-ranked set of documents post-RLM feedback, the RLM being estimated on $R_M(Q)$ - the top-$M$ documents, and $\xi$ is a rank correlation coefficient of two ordered sets, for which we specifically use \pr, as suggested in \cite{uef_kurland_sigir10}.
We experiment with three specific instances of the base estimators, namely Clarity, WIG and NQC for UEF, which we denote as UEF(Clarity), UEF(WIG) and UEF(NQC), respectively.

\end{compactitem}

\subsubsection{Parameters and settings.}
The standard practice in QPP research is to optimize the common hyper-parameter - the number of top documents of post-retrieval QPP approaches (denoted as $k$ in Definition \ref{def:qppmethod}). This hyper-parameter is tuned via a grid search on a development set of queries and the optimal setting is used to report the performance on a test set. A common approach is to employ a 50:50 split of the set of queries into development and test sets. This process is usually repeated 30 times and the average results over the test folds are reported \cite{kurland_tois12,query_variants_kurland,wig_croft_SIGIR07}.

The focus of our research is different, however, in the sense that we seek to analyze the variations caused due to different settings for constructing the QPP ground-truth, instead of demonstrating that a particular QPP method outperforms others. Moreover, an optimal tuning of the hyper-parameters for each QPP method would require averaging over 30 different experiments for a single way of defining the QPP context for constructing the ground-truth. Hence, to keep the number of experiments tractable, we set $k=20$, as frequently prescribed in the literature \cite{croft_qpp_sigir02,kurland_tois12,query_variants_kurland,wig_croft_SIGIR07}. Another hyper-parameter, specific to UEF, is the number of times a subset of size $k$ is sampled from a set of top-$K$ ($K > k$) documents. We use a total of $10$ random samples of $k=20$ documents from the set of $K=100$ top documents, as prescribed in \cite{uef_kurland_sigir10}.

\begin{table}[!ht]
\begin{subtable}{.5\linewidth}
\centering
\begin{adjustbox}{width=\textwidth}
\begin{tabular}{@{}llccccr@{}}

\toprule
 & & \multicolumn{4}{c}{IR Evaluation Metric ($\theta$)} &  
\\ 

\cmidrule{3-6}

& Model($\mathcal{S}$) & AP & nDCG & R & P@10 & $\sigma({\theta})$
\\

\midrule

 
 & \lmjm & 0.3795 & 0.3966 & 0.3869 & 0.3311 & \cellcolor{magicmint}\textbf{0.0291} \\
 \textbf{$r$} & BM25 & 0.5006 & 0.4879 & 0.4813 & 0.2525 & \cellcolor{melon}\textbf{0.1190} \\
 & \lmdr & 0.5208 & 0.5062 & 0.4989 & 0.2851 & 0.1121 \\
\cmidrule{3-6}
 & $\sigma({\mathcal{S}})$ & \cellcolor{melon}\textbf{0.0764} & 0.0587 & 0.0602 & \cellcolor{magicmint}0.0395 \\

\midrule

 
 & \lmjm & 0.4553 & 0.4697 & 0.4663 & 0.3067 & \cellcolor{magicmint}0.0788 \\
 \textbf{$\rho$} & BM25 & 0.4526 & 0.4700 & 0.4736&  0.2842 & \cellcolor{melon}0.0911 \\
 & \lmdr & 0.4695 & 0.4848 & 0.4893 & 0.3017 & 0.0902 \\
\cmidrule{3-6}
 & $\sigma({\mathcal{S}})$ & 0.0091 & \cellcolor{magicmint}0.0086 & \cellcolor{melon}0.0118 & 0.0114 \\

\midrule

 
 & \lmjm & 0.3175 & 0.3285 & 0.3278 & 0.2193 & \cellcolor{magicmint}0.0529 \\
 \textbf{$\tau$} & BM25 & 0.3144 & 0.3162 & 0.3319 & 0.2040 & 0.0589 \\
 & \lmdr & 0.3307 & 0.3407 & 0.3440 & 0.2155 & \cellcolor{melon}0.0617 \\
\cmidrule{3-6}
 & $\sigma({\mathcal{S}})$ & 0.0087 & \cellcolor{melon}0.0123 & \cellcolor{magicmint}\textbf{0.0084} & 0.0120 \\

\bottomrule
\end{tabular}

\end{adjustbox}
\caption{\small AvgIDF}
\end{subtable}%
\quad
\begin{subtable}{.5\linewidth}
\centering
\begin{adjustbox}{width=\textwidth}
\begin{tabular}{@{}llccccc@{}}

\toprule
 & & \multicolumn{4}{c}{IR Evaluation Metric ($\theta$)} &  
\\ 

\cmidrule{3-6}

& Model($\mathcal{S}$) & AP & nDCG & R & P@10 & $\sigma({\theta})$ \\

\midrule

 
 & \lmjm & 0.3652 & 0.4169 & 0.4503 & 0.2548 & 0.0855 \\
 \textbf{$r$} & BM25 & 0.3563 & 0.4118 & 0.4495 & 0.2707 & \cellcolor{magicmint}0.0777 \\
 & \lmdr & 0.4354 & 0.4583 & 0.4854 & 0.2842 & \cellcolor{melon}0.0901 \\
\cmidrule{3-6}
 & $\sigma({\mathcal{S}})$ & \cellcolor{melon}\textbf{0.0433} & 0.0255 & 0.0205 & \cellcolor{magicmint}0.0147 \\

\midrule

 
 & \lmjm & 0.4545 & 0.4843 & 0.5248 & 0.2918 & \cellcolor{melon}\textbf{0.1022} \\
 \textbf{$\rho$} & BM25 & 0.4618 & 0.4887 & 0.5137 & 0.3308 & \cellcolor{magicmint}0.0814 \\
 & \lmdr & 0.5024 & 0.5260 & 0.5453 & 0.3340 & 0.0969 \\
\cmidrule{3-6}
 & $\sigma({\mathcal{S}})$ & \cellcolor{melon}0.0258 & 0.0229 & \cellcolor{magicmint}0.0160 & 0.0235 \\

\midrule

 
 & \lmjm & 0.3100 & 0.3319 & 0.3657 & 0.2061 & \cellcolor{melon}0.0688 \\
 \textbf{$\tau$} & BM25 & 0.3170 & 0.3370 & 0.3551 & 0.2374 & \cellcolor{magicmint}\textbf{0.0519} \\
 & \lmdr & 0.3539 & 0.3713 & 0.3828 & 0.2379 & 0.0668 \\
\cmidrule{3-6}
 & $\sigma({\mathcal{S}})$ & \cellcolor{melon}0.0236 & 0.0214 & \cellcolor{magicmint}\textbf{0.0140} & 0.0182 \\

\bottomrule
\end{tabular}

\end{adjustbox}
\caption{\small NQC}
\end{subtable}
\begin{subtable}{.5\linewidth}
\centering
\begin{adjustbox}{width=\textwidth}
\begin{tabular}{@{}llccccc@{}}

\toprule
 & & \multicolumn{4}{c}{IR Evaluation Metric ($\theta$)} &  
\\ 

\cmidrule{3-6}

& Model($\mathcal{S}$) & AP & nDCG & R & P@10 & $\sigma({\theta})$
\\

\midrule


& \lmjm & 0.4056 & 0.4071 & 0.3971 & 0.3054 & \cellcolor{magicmint}0.0491 \\
\textbf{$r$} & BM25 & 0.4488 & 0.4563 & 0.4386 & 0.3485 & 0.0502 \\
& \lmdr & 0.4908 & 0.4798 & 0.4632 & 0.3423 & \cellcolor{melon}\textbf{0.0688} \\
\cmidrule{3-6}
& $\sigma({\mathcal{S}})$ & \cellcolor{melon}0.0426 & 0.0371 & 0.0334 & \cellcolor{magicmint}0.0233 \\

\midrule


& \lmjm & 0.3716 & 0.3794 & 0.3790 & 0.3120 & \cellcolor{magicmint}0.0325 \\
\textbf{$\rho$} & BM25 & 0.4520 & 0.4601 & 0.4505 & 0.3586 & 0.0480 \\
& \lmdr & 0.4582 & 0.4688 & 0.4667 & 0.3528 & \cellcolor{melon}0.0561 \\
\cmidrule{3-6}
& $\sigma({\mathcal{S}})$ & 0.0483 & \cellcolor{melon}\textbf{0.0493} & 0.0467 & \cellcolor{magicmint}0.0254 \\

\midrule


& \lmjm & 0.2514 & 0.2567 & 0.2607 & 0.2209 & \cellcolor{magicmint}\textbf{0.0181} \\
\textbf{$\tau$} & BM25 & 0.3116 & 0.3181 & 0.3125 & 0.2549 & 0.0297 \\
& \lmdr & 0.3194 & 0.3267 & 0.3259 & 0.2493 & \cellcolor{melon}0.0375 \\
\cmidrule{3-6}
& $\sigma({\mathcal{S}})$ & 0.0372 & \cellcolor{melon}0.0382 & 0.0344 & \cellcolor{magicmint}\textbf{0.0182} \\

\bottomrule
\end{tabular}

\end{adjustbox}
\caption{\small WIG}
\end{subtable}%
\quad
\begin{subtable}{.5\linewidth}
\centering
\begin{adjustbox}{width=\textwidth}
\begin{tabular}{@{}llccccc@{}}

\toprule
 & & \multicolumn{4}{c}{IR Evaluation Metric ($\theta$)} &  
\\ 

\cmidrule{3-6}

& Model($\mathcal{S}$) & AP & nDCG & R & P@10  & $\sigma({\theta})$ 
\\

\midrule


& \lmjm & 0.4746 & 0.4763 & 0.4646 & 0.3573 & \cellcolor{magicmint}0.0575 \\
\textbf{$r$} & BM25 & 0.5386 & 0.5476 & 0.5263 & 0.4182 & 0.0603 \\
& \lmdr & 0.5693 & 0.5566 & 0.5373 & 0.3971 & \cellcolor{melon}\textbf{0.0797} \\
\cmidrule{3-7}
 & $\sigma({\mathcal{S}})$ & \cellcolor{melon}0.0483 & 0.0440 & 0.0392 & \cellcolor{magicmint}0.0309 \\

\midrule


& \lmjm & 0.4385 & 0.4477 & 0.4472 & 0.3682 & \cellcolor{magicmint}0.0384 \\
\textbf{$\rho$} & BM25 & 0.5334 & 0.5429 & 0.5316 & 0.4231 & 0.0567 \\
& \lmdr & 0.5407 & 0.5532 & 0.5507 & 0.4163 & \cellcolor{melon}0.0662 \\
\cmidrule{3-6}
& $\sigma({\mathcal{S}})$ & 0.0570 & \cellcolor{melon}\textbf{0.0582} & 0.0551 & \cellcolor{magicmint}0.0300 \\

\midrule


& \lmjm & 0.3017 & 0.3080 & 0.3128 & 0.2651 & \cellcolor{magicmint}\textbf{0.0217} \\
\textbf{$\tau$} & BM25 & 0.3677 & 0.3754 & 0.3688 & 0.3008 & 0.0351 \\
& \lmdr & 0.3833 & 0.3920 & 0.3911 & 0.2992 & \cellcolor{melon}0.0450 \\
\cmidrule{3-6}
& $\sigma({\mathcal{S}})$ & 0.0433 & \cellcolor{melon}0.0445 & 0.0303 & \cellcolor{magicmint}\textbf{0.0202} \\

\bottomrule
\end{tabular}

\end{adjustbox}
\caption{\small UEF(WIG)}
\end{subtable}
\vskip 0.5em
\caption{
\small
Sensitivity of QPP results with variations in the IR evaluation metric ($\theta$) and the IR model ($\mathcal{S}$) for the QPP methods a) AvgIDF, b) NQC, c) WIG and d) UEF(WIG). The metrics - AP, nDCG and recall (R) are measured on the top-100 retrieved documents using retrieval models \lmjm($\lambda=0.6$), BM$25(k_1=0.7,b=0.3)$ and \lmdr($\mu=1000$) respectively.
The lowest (highest) standard deviations for each group of QPP correlation measure are shown in green (red). The lowest and the highest across different correlation measures are shown bold-faced.
\label{tab:sdres}
}
\end{table}
\begin{table}[ht]
\centering
\begin{adjustbox}{width=.9\textwidth}

\begin{tabular}{@{}lccccccccc@{}}

\toprule

Model & Metric & 
AP@100 & AP@1000 & 
R@10 & R@100 & R@1000 & 
nDCG@10 & nDCG@100 & nDCG@1000 \\

\midrule

\lmjm & \multirow{3}{*}{AP@10} & 
0.4286 & 0.3333 & 
0.9048 & 0.2381 & \cellcolor{melon}\textbf{-0.1429} &
1.0000 & 0.2381 & 0.3333 \\

BM25 & & 
1.0000 & 0.9048 & 
1.0000 & 0.9048 & 0.4286 &
1.0000 & 1.0000 & 0.7143 \\

\lmdr & & 
1.0000 & 0.9048 & 
1.0000 & 0.9048 & 0.4286 &
1.0000 & 1.0000 & 0.7143 \\

\cmidrule{1-2}

\lmjm & \multirow{3}{*}{AP@100} & 
\nores & 0.9048 & 
0.5238 & 0.8095 & \cellcolor{melon}0.4286 &
\cellcolor{melon}0.4286 & 0.8095 & 0.9048 \\

BM25 & & 
\nores & 0.9048 & 
1.0000 & 0.9048 & \cellcolor{melon}0.4286 &
1.0000 & 1.0000 & 0.7143 \\

\lmdr & & 
\nores & 0.9048 & 
1.0000 & 0.9048 & \cellcolor{melon}0.4286 &
1.0000 & 1.0000 & 0.7143 \\

\cmidrule{1-2}

\lmjm & \multirow{3}{*}{AP@1000} & 
\nores & \nores & 
0.4286 & 0.8095 & 0.5238 &
\cellcolor{melon}0.3333 & 0.9048 & 1.0000 \\

BM25 & & 
\nores & \nores & 
0.9048 & 0.8095 & \cellcolor{melon}0.3333 &
0.9048 & 0.9048 & 0.8095 \\

\lmdr & & 
\nores & \nores & 
0.9048 & 0.8095 & 0.5238 &
0.9048 & 0.9048 & 0.8095 \\

\cmidrule{1-2}

\lmjm & \multirow{3}{*}{R@10} & 
\nores & \nores & 
\nores & 0.3333 & \cellcolor{melon}-0.0476 &
0.9048 & 0.3333 & 0.4286 \\

BM25 & & 
\nores & \nores & 
\nores & 0.9048 & 0.4286 &
1.0000 & 1.0000 & 0.7143 \\

\lmdr & & 
\nores & \nores & 
\nores & 0.9048 & 0.4286 &
1.0000 & 1.0000 & 0.7143 \\

\cmidrule{1-2}

\lmjm & \multirow{3}{*}{R@100} & 
\nores & \nores & 
\nores & \nores & 0.6190 &
\cellcolor{melon}0.2381 & 1.0000 & 0.9048 \\

BM25 & & 
\nores & \nores & 
\nores & \nores & 0.5238 &
0.9048 & 0.9048 & 0.6190 \\

\lmdr & & 
\nores & \nores & 
\nores & \nores & 0.5238 &
0.9048 & 0.9048 & 0.6190 \\

\cmidrule{1-2}

\lmjm & \multirow{3}{*}{R@1000} & 
\nores & \nores & 
\nores & \nores & \nores &
\cellcolor{melon}\textbf{-0.1429} & 0.6190 & 0.5238 \\ 

BM25 & & 
\nores & \nores & 
\nores & \nores & \nores &
0.4286 & 0.4286 & 0.5238 \\

\lmdr & & 
\nores & \nores & 
\nores & \nores & \nores &
0.4286 & 0.4286 & 0.5238 \\

\cmidrule{1-2}

\lmjm & \multirow{3}{*}{nDCG@10} & 
\nores & \nores & 
\nores & \nores & \nores &
\nores & \cellcolor{melon}0.2381 & 0.3333 \\

BM25 & & 
\nores & \nores & 
\nores & \nores & \nores &
\nores & 1.0000 & 0.7143 \\

\lmdr & & 
\nores & \nores & 
\nores & \nores & \nores &
\nores & 1.0000 & 0.7143 \\

\cmidrule{1-2}

\lmjm & \multirow{3}{*}{nDCG@100} & 
\nores & \nores & 
\nores & \nores & \nores &
\nores & \nores & \cellcolor{melon}0.9048 \\

BM25 & & 
\nores & \nores & 
\nores & \nores & \nores &
\nores & \nores & 0.7143 \\

\lmdr & & 
\nores & \nores & 
\nores & \nores & \nores &
\nores & \nores & 0.7143 \\

\bottomrule
\end{tabular}
\end{adjustbox}
\vskip 0.5em
\caption{
\small
Each cell in the table indicates the correlation (Kendall's $\tau$) between QPP systems ranked in order by their evaluated effectiveness (measured with the help of Pearson's $r$ for the results of this table) for two different IR metrics corresponding to the row and the column name of the cell. A total of $7$ QPP systems were used in these experiments, namely AvgIDF, Clarity, WIG, NQC, UEF(Clarity), UEF(WIG) and UEF(NQC). The lowest correlation value for each group is marked in red, and the lowest correlations, overall, are bold-faced.
}
\label{tab:cont_met_r}
\end{table}
\begin{table}[!ht]
\centering
\begin{adjustbox}{width=.9\textwidth}

\begin{tabular}{@{}lccccccccc@{}}

\toprule

Model & Metric & 
AP@100 & AP@1000 & 
R@10 & R@100 & R@1000 & 
nDCG@10 & nDCG@100 & nDCG@1000 \\

\midrule

\lmjm & \multirow{3}{*}{AP@10} & 
0.5238 & 0.3333 & 
0.8095 & 0.4286 & \cellcolor{melon}0.2381 &
0.8095 & 0.4286 & 0.3333 \\

BM25 & & 
0.9048 & 0.7143 & 
0.8095 & 0.8095 & 0.5238 &
1.0000 & 0.9048 & 0.5238 \\

\lmdr & & 
0.9048 & 0.8095 & 
1.0000 & 1.0000 & 0.8095 &
1.0000 & 0.9048 & 0.7143 \\

\cmidrule{1-2}

\lmjm & \multirow{3}{*}{AP@100} & 
\nores & 0.8095 & 
0.5238 & 0.9048 & 0.7143 &
\cellcolor{melon}0.3333 & 0.9048 & 0.8095 \\

BM25 & & 
\nores & 0.8095 & 
0.9048 & 0.9048 & 0.6190 &
0.9048 & 1.0000 & 0.6190 \\

\lmdr & & 
\nores & 0.9048 & 
0.9048 & 0.9048 & 0.7143 &
0.9048 & 1.0000 & 0.8095 \\

\cmidrule{1-2}

\lmjm & \multirow{3}{*}{AP@1000} & 
\nores & \nores & 
0.3333 & 0.9048 & 0.7143 &
\cellcolor{melon}\textbf{0.1429} & 0.9048 & 1.0000 \\

BM25 & & 
\nores & \nores & 
0.7143 & 0.7143 & 0.6190 &
0.7143 & 0.8095 & 0.8095 \\

\lmdr & & 
\nores & \nores & 
0.8095 & 0.8095 & 0.8095 &
0.8095 & 0.9048 & 0.9048 \\

\cmidrule{1-2}

\lmjm & \multirow{3}{*}{R@10} & 
\nores & \nores & 
\nores & 0.4286 & \cellcolor{melon}0.2381 &
0.8095 & 0.4286 & 0.3333 \\

BM25 & & 
\nores & \nores & 
\nores & 1.0000 & 0.7143 &
0.8095 & 1.0000 & 0.5238 \\

\lmdr & & 
\nores & \nores & 
\nores & 1.0000 & 0.8095 &
1.0000 & 0.9048 & 0.7143 \\

\cmidrule{1-2}

\lmjm & \multirow{3}{*}{R@100} & 
\nores & \nores & 
\nores & \nores & 0.8095 &
\cellcolor{melon}0.2381 & 1.0000 & 0.9048 \\

BM25 & & 
\nores & \nores & 
\nores & \nores & 0.7143 &
0.8095 & 0.9048 & 0.5238 \\

\lmdr & & 
\nores & \nores & 
\nores & \nores & 0.8095 &
1.0000 & 0.9048 & 0.7143 \\

\cmidrule{1-2}

\lmjm & \multirow{3}{*}{R@1000} & 
\nores & \nores & 
\nores & \nores & \nores &
\cellcolor{melon}0.0476 & 0.8095 & 0.7143 \\

BM25 & & 
\nores & \nores & 
\nores & \nores & \nores &
0.5238 & 0.6190 & 0.8095 \\

\lmdr & & 
\nores & \nores & 
\nores & \nores & \nores &
0.8095 & 0.7143 & 0.9048 \\

\cmidrule{1-2}

\lmjm & \multirow{3}{*}{nDCG@10} & 
\nores & \nores & 
\nores & \nores & \nores &
\nores & 0.2381 & \cellcolor{melon}\textbf{0.1429} \\

BM25 & & 
\nores & \nores & 
\nores & \nores & \nores &
\nores & 0.9048 & 0.5238 \\

\lmdr & & 
\nores & \nores & 
\nores & \nores & \nores &
\nores & 0.9048 & 0.7143 \\

\cmidrule{1-2}

\lmjm & \multirow{3}{*}{nDCG@100} & 
\nores & \nores & 
\nores & \nores & \nores &
\nores & \nores & 0.9048 \\

BM25 & & 
\nores & \nores & 
\nores & \nores & \nores &
\nores & \nores & \cellcolor{melon}0.6190 \\

\lmdr & & 
\nores & \nores & 
\nores & \nores & \nores &
\nores & \nores & 0.8095 \\

\bottomrule
\end{tabular}
\end{adjustbox}
\vskip 0.5em
\caption{
\small
Results of relative changes in the ranks of QPP systems (similar to Table \ref{tab:cont_met_r}), the difference being that the QPP outcomes were measured with $\tau$ (instead of $r$).
}
\label{tab:cont_met_tau}
\end{table}
\begin{table}[!ht]
\centering
\small
\begin{adjustbox}{width=0.83\textwidth}

\begin{tabular}{lcccccccc}

\toprule

Metric & Model & 
\lmjm &
BM25 & BM25 & BM25 & 
\lmdr & \lmdr & \lmdr \\

& & ($0.6$) &
($0.7, 0.3$) & ($1.0, 1.0$) & ($0.3, 0.7$) &
($100$) & ($500$) & ($1000$) \\

\midrule

AP@100 & & 
1.0000 & 
0.9048 & 1.0000 & 0.9048 & 
0.9048 & 0.9048 & 0.9048 \\

nDCG@100 & \lmjm & 
1.0000 & 
0.8095 & 0.9048 & 0.9048 & 
0.9048 & 0.8095 & 0.8095 \\

R@100 & ($0.3$) & 
0.9048 & 
0.8095 & 0.9048 & 1.0000 & 
1.0000 & 0.9048 & 0.9048 \\

P@10 & & 
1.0000 & 
0.8095 & 1.0000 & 0.8095 & 
\cellcolor{melon}\textbf{0.7143} & \cellcolor{melon}\textbf{0.7143} & 1.0000 \\

\cmidrule{1-2}

AP@100 & &  
\nores & 
0.9048 & 1.0000 & 0.9048 & 
0.9048 & 0.9048 & 0.9048 \\

nDCG@100 & \lmjm & 
\nores & 
0.8095 & 0.9048 & 0.9048 & 
0.9048 & 0.8095 & 0.8095 \\

R@100 & ($0.6$) & 
\nores & 
0.9048 & 1.0000 & 0.9048 & 
0.9048 & 1.0000 & 1.0000 \\

P@10 & & 
\nores & 
0.8095 & 1.0000 & 0.8095 & 
\cellcolor{melon}\textbf{0.7143} & \cellcolor{melon}\textbf{0.7143} & 1.0000 \\

\cmidrule{1-2}

AP@100 & &
\nores & 
\nores & 0.9048 & 0.9048 & 
1.0000 & 1.0000 & 1.0000 \\

nDCG@100 & BM25 &
\nores & 
\nores & 0.9048 & 0.9048 & 
0.9048 & 1.0000 & 1.0000 \\

R@100 & ($0.7, 0.3$) &
\nores & 
\nores & 0.9048 & \cellcolor{melon}0.8095 & 
\cellcolor{melon}0.8095 & 0.9048 & 0.9048 \\

P@10 & & 
\nores & 
\nores & \cellcolor{melon}0.8095 & 1.0000 & 
0.9048 & 0.9048 & \cellcolor{melon}0.8095 \\

\cmidrule{1-2}

AP@100 & & 
\nores & 
\nores & \nores & 0.9048 & 
0.9048 & 0.9048 & 0.9048 \\

nDCG@100 & BM25 &
\nores & 
\nores & \nores & 1.0000 & 
1.0000 & 0.9048 & 0.9048 \\

R@100 & ($1.0, 1.0$) & 
\nores & 
\nores & \nores & 0.9048 & 
0.9048 & 1.0000 & 1.0000 \\

P@10 & & 
\nores & 
\nores & \nores & 0.8095 & 
\cellcolor{melon}\textbf{0.7143} & \cellcolor{melon}\textbf{0.7143} & 1.0000 \\

\cmidrule{1-2}

AP@100 & &  
\nores & 
\nores & \nores & \nores & 
1.0000 & 1.0000 & 1.0000 \\

nDCG@100 & BM25 & 
\nores & 
\nores & \nores & \nores & 
1.0000 & 0.9048 & 0.9048 \\

R@100 & ($0.3, 0.7$) & 
\nores & 
\nores & \nores & \nores & 
1.0000 & 0.9048 & 0.9048 \\

P@10 & & 
\nores & 
\nores & \nores & \nores & 
0.9048 & 0.9048 & \cellcolor{melon}0.8095 \\

\cmidrule{1-2}

AP@100 & & 
\nores & 
\nores & \nores & \nores & 
\nores & 1.0000 & 1.0000 \\

nDCG@100 & \lmdr &
\nores & 
\nores & \nores & \nores & 
\nores & 0.9048 & 0.9048 \\

R@100 & ($100$) & 
\nores & 
\nores & \nores & \nores & 
\nores & 0.9048 & 0.9048 \\

P@10 & &  
\nores & 
\nores & \nores & \nores & 
\nores & 0.8095 & \cellcolor{melon}\textbf{0.7143} \\

\cmidrule{1-2}

AP@100 & &  
\nores & 
\nores & \nores & \nores & 
\nores & \nores & 1.0000 \\

nDCG@100 & \lmdr & 
\nores & 
\nores & \nores & \nores & 
\nores & \nores & 1.0000 \\

R@100 & ($500$) &  
\nores & 
\nores & \nores & \nores & 
\nores & \nores & 1.0000 \\

P@10 & & 
\nores & 
\nores & \nores & \nores & 
\nores & \nores & \cellcolor{melon}\textbf{0.7143} \\

\bottomrule
\end{tabular}
\end{adjustbox}
\vskip 0.5em
\caption{
\small
Each cell in the table indicates the correlation (Kendall's $\tau$) between QPP systems ranked in order by their evaluated effectiveness (measured with the help of Pearson's $r$ for the results presented in this table) for each pair of IR models for $7$ different QPP systems.
The lowest correlation value for each group is marked in red. The lowest correlation in the table is bold-faced.
%
}
\label{tab:cont_model_r}
\end{table}
\begin{table}[!ht]
\centering
\begin{adjustbox}{width=0.83\textwidth}

\begin{tabular}{lcccccccc}

\toprule


Metric & Model & 
\lmjm &
BM25 & BM25 & BM25 & 
\lmdr & \lmdr & \lmdr \\

& & ($0.6$) &
($0.7, 0.3$) & ($1.0, 1.0$) & ($0.3, 0.7$) &
($100$) & ($500$) & ($1000$) \\

\midrule

AP@100 & & 
1.0000 & 
1.0000 & 1.0000 & 1.0000 & 
1.0000 & 1.0000 & 1.0000 \\

nDCG@100 & \lmjm & 
1.0000 & 
1.0000 & 1.0000 & 1.0000 & 
1.0000 & 1.0000 & 1.0000 \\

R@100 & ($0.3$) & 
1.0000 & 
1.0000 & 1.0000 & 1.0000 & 
1.0000 & 1.0000 & 1.0000 \\

P@10 & & 
0.9048 & 
1.0000 & 0.9048 & \cellcolor{melon}0.8095 & 
0.9095 & 1.0000 & 1.0000 \\

\cmidrule{1-2}

AP@100 & &
\nores & 
1.0000 & 1.0000 & 1.0000 & 
1.0000 & 1.0000 & 1.0000 \\

nDCG@100 & \lmjm &
\nores & 
1.0000 & 1.0000 & 1.0000 & 
1.0000 & 1.0000 & 1.0000 \\

R@100 & ($0.6$) &  
\nores & 
1.0000 & 1.0000 & 1.0000 & 
1.0000 & 1.0000 & 1.0000 \\

P@10 & & 
\nores & 
0.9048 & 1.0000 & \cellcolor{melon}0.7143 & 
\cellcolor{melon}0.7143 & 0.9048 & 0.9048 \\

\cmidrule{1-2}

AP@100 & &
\nores & 
\nores & 1.0000 & 1.0000 & 
1.0000 & 1.0000 & 1.0000 \\

nDCG@100 & BM25 & 
\nores & 
\nores & 1.0000 & 1.0000 & 
1.0000 & 1.0000 & 1.0000 \\

R@100 & ($0.7, 0.3$) & 
\nores & 
\nores & 1.0000 & 1.0000 & 
1.0000 & 1.0000 & 1.0000 \\

P@10 & & 
\nores & 
\nores & 0.9048 & \cellcolor{melon}0.8095 & 
\cellcolor{melon}0.8095 & 1.0000 & 1.0000 \\

\cmidrule{1-2}

AP@100 & & 
\nores & 
\nores & \nores & 1.0000 & 
1.0000 & 1.0000 & 1.0000 \\

nDCG@100 & BM25 & 
\nores & 
\nores & \nores & 1.0000 & 
1.0000 & 1.0000 & 1.0000 \\

R@100 & ($1.0, 1.0$) & 
\nores & 
\nores & \nores & 1.0000 & 
1.0000 & 1.0000 & 1.0000 \\

P@10 & & 
\nores & 
\nores & \nores & \cellcolor{melon}0.7143 & 
\cellcolor{melon}0.7143 & 0.9048 & 0.9048 \\

\cmidrule{1-2}

AP@100 & &
\nores & 
\nores & \nores & \nores & 
1.0000 & 1.0000 & 1.0000 \\

nDCG@100 & BM25 & 
\nores & 
\nores & \nores & \nores & 
1.0000 & 1.0000 & 1.0000 \\

R@100 & ($0.3, 0.7$) &  
\nores & 
\nores & \nores & \nores & 
1.0000 & 1.0000 & 1.0000 \\

P@10 & & 
\nores & 
\nores & \nores & \nores & 
\cellcolor{melon}\textbf{0.6190} & 0.8095 & 0.8095 \\

\cmidrule{1-2}

AP@100 & &
\nores & 
\nores & \nores & \nores & 
\nores & 1.0000 & 1.0000 \\

nDCG@100 & \lmdr & 
\nores & 
\nores & \nores & \nores & 
\nores & 1.0000 & 1.0000 \\

R@100 & ($100$) &  
\nores & 
\nores & \nores & \nores & 
\nores & 1.0000 & 1.0000 \\

P@10 & & 
\nores & 
\nores & \nores & \nores & 
\nores & \cellcolor{melon}0.8095 & \cellcolor{melon}0.8095 \\

\cmidrule{1-2}

AP@100 & & 
\nores & 
\nores & \nores & \nores & 
\nores & \nores & 1.0000 \\

nDCG@100 & \lmdr & 
\nores & 
\nores & \nores & \nores & 
\nores & \nores & 1.0000 \\

R@100 & ($500$) & 
\nores & 
\nores & \nores & \nores & 
\nores & \nores & 1.0000 \\

P@10 & & 
\nores & 
\nores & \nores & \nores & 
\nores & \nores & 1.0000 \\

\bottomrule
\end{tabular}
\end{adjustbox}
\vskip 0.5em
\caption{
\small
The difference of this table with Table \ref{tab:cont_model_r} is that the QPP effectiveness is measured with Kendall's $\tau$ (instead of Pearson's $r$ as in Table \ref{tab:cont_model_r}).
}
\label{tab:cont_model_tau}
\end{table}

\section{Results}
\label{sec:res}
\subsection{RQ1: Variations in QPP Evaluations}
Table \ref{tab:sdres} reports the standard deviations in the observed values for the QPP experiments\footnote{Tables \ref{tab:sdres}-\ref{tab:cont_model_tau} are best viewed in color.}.
In Tables \ref{tab:sdres}a-d,
the value of $\sigma(\theta)$ in each row indicates the standard deviation of the QPP outcome values observed in that row, i.e., these values indicate the standard deviation resulting from the use of different IR metrics for QPP evaluation. Similarly, the value of $\sigma(\mathcal{S})$ in each column is the standard deviation of the $r$, $\rho$ or $\tau$ values reported in that column, i.e., this value denotes the standard deviations in QPP correlations across different IR models. 
The lowest standard deviations for each QPP correlation type are shown bold-faced.
We now discuss the observations that can be made from 
Table \ref{tab:sdres}.


\para{Variations due to IR evaluation metric}
The first set of observations, listed below, is in relation to the absolute differences between two different QPP evaluations involving two different QPP contexts.
\begin{compactitem}
\item \textbf{Substantial absolute differences in the QPP outcomes}: 
Variations in the IR evaluation metric (i.e., the $\theta$ component of a QPP context $\mathcal{C}(Q)$ of Definition \ref{def:context}) while keeping the other two components fixed (i.e., retrieval model and cut-off) yields considerable absolute differences in the values. As an example, compare the QPP evaluation of $0.5006$ with AP@100 in Table \ref{tab:sdres}a to that of $0.2525$ with P@10 obtained with BM25, showing that these absolute differences can be high.

\item \textbf{Lower variations with $\tau$}: In general, we observe that each QPP method (e.g. NQC, WIG etc.) exhibits considerable differences in measured outcomes specially between AP@100 and P@10. Moreover, the variations, in general, are lower when correlation is measured with the help of Kendall's $\tau$ (e.g., compare $\sigma(\theta)=0.0181$ measured with $\tau$ vs. $\sigma(\theta)=0.0491$ measured with $r$ on documents retrieved with \lmjm).
The fact that $\tau$ exhibits a lower variance in QPP evaluation is likely because the correlation is measured in a pairwise manner ($\tau$ being a function of the number of concordant and discordant pairs). As a result, $\tau$ depends only on the agreements between the true and the predicted order (of query difficulty) between a query pair, and not on the absolute values of the predicted scores or the reference values of the IR evaluation metric (as in Pearson's $r$ or Spearman's $\rho$). 

\item \textbf{Lower variances with \lmjm}: Similar to our earlier observation that $\tau$ should be the preferred QPP evaluation measure (with an objective to minimize the variances in observed results due to changes in IR evaluation metric), we observe from Table \ref{tab:sdres} that \lmjm, in most cases, result in low variances in QPP experiment outcomes.  

\end{compactitem}

\para{Variations due to IR models}
The second set of observations from Table \ref{tab:sdres} relates to variations in the observed QPP results with respect to variations in IR models. The standard deviations of these values correspond to column-wise calculation of standard deviations and are shown as the $\sigma(\mathcal{S})$ values. Again, similar to the $\sigma(\theta)$ values, the lowest (highest) values along each row of $\sigma(\mathcal{S})$ are colored in green (red) to reflect the situation of lower the better. The best values across different QPP correlations are bold-faced. We summarise our observations:
\begin{compactitem}
\item \textbf{Lower variations with $\tau$}: Similar to the $\sigma(\theta)$ values it is again observed that mostly measuring QPP outcomes with $\tau$ results in the lowest variances in QPP results. Consequently, for better reproducibility it is more useful to report results with Kendall's $\tau$.

\item \textbf{Lower variations in the QPP outcomes}: Compared to variations across IR evaluation metrics, we observe that the variations occurring across IR models is lower (compare the bold-faced green $\sigma(\mathcal{S})$ values with those of $\sigma(\theta)$ ones). This entails that experiments need to put more emphasis on a precise description of the IR metrics used for QPP evaluation.

\item \textbf{Lack of a consistency on which combination of QPP method with IR evaluation context yields least the variance}:
While WIG and UEF(WIG) exhibit lowest variances for a precision oriented evaluation of ground-truth retrieval effectiveness, for AvgIDF and NQC methods, the least variations are noted for recall.
\end{compactitem}

\subsection{RQ2: Variations in the Relative Ranks of QPP Methods}

We now report results in relation to the second research question, where the intention is to measure how stable are QPP system ranks (ordered by their evaluated effectiveness measures) for variations in the QPP context.

\para{Variation due to IR metrics}
Tables \ref{tab:cont_met_r} and \ref{tab:cont_met_tau} present the pairwise contingency table for different combinations of IR metrics for three different IR models. The following observations can be made from the results.

\begin{compactitem}
\item \textbf{\lmjm~leads to the most instability in the relative QPP system ranks}: This behaviour, most likely, can be attributed to the fact that this model has a tendency to favour shorter documents in the top-retrieved in contrast to \lmdr~or BM25.
\item \textbf{Some evaluation metrics are more sensitive to rank cut-off values}: For instance, the QPP ground-truth measured with Recall@10 yields considerably different results when the ground-truth corresponds to Recall@1000.

\item \textbf{Relative ranks of QPP systems more stable with $\tau$}: A comparison between the values of Tables \ref{tab:cont_met_r} and \ref{tab:cont_met_tau} reveals that a rank correlation measure such as $\tau$ leads to better stability of QPP experiments than when $r$ is used to measure the relative effectiveness of QPP models.   
\end{compactitem}

\para{Variations due to IR models}
Tables \ref{tab:cont_model_r} and \ref{tab:cont_model_tau} present the pairwise contingency between retrieval similarity scores from different evaluation metrics.
For this set of experiments, the intention is also to investigate the stability of QPP system ranks with respect to changes, not only to the retrieval model itself, but also for different parameter settings on the same model, e.g. BM25(0.7,0.3)\footnote{Values of $k_1$ and $b$, respectively, in BM25 \cite{Okapi}.} vs. BM25(1,1). We observe the following:
\begin{compactitem}
\item \textbf{Relative ranks of QPP systems are quite stable across IR models}: The correlation values of Tables \ref{tab:cont_model_r} and \ref{tab:cont_model_tau} are higher than those of Tables \ref{tab:cont_met_r} and \ref{tab:cont_met_tau}, which shows that the QPP experiments are less sensitive to  variations in the set of top documents retrieved by different similarity scores.

\item \textbf{\lmjm~leads to more instability in the QPP outcomes}: \lmjm~shows the lowest correlation with other retrieval models. Parameter variations of an IR model usually lead to relatively stable QPP outcomes. For instance, see the correlations between \lmdr(500) and \lmdr(1000).

\item \textbf{Relative ranks of QPP systems are more stable with $\tau$}: This observation (a comparison between the values of Tables \ref{tab:cont_model_r} and \ref{tab:cont_model_tau}) is similar to the comparison between Tables \ref{tab:cont_met_r} and \ref{tab:cont_met_tau}. However, the differences between the correlation values are smaller in comparison to those observed between Tables \ref{tab:cont_met_r} and \ref{tab:cont_met_tau}.
\end{compactitem}

\section{Concluding Remarks}
 \label{sec:conclusion}
We have shown via extensive experiments that QPP outcomes are indeed sensitive to the experimental configuration used. As part of our analysis, we have found that certain factors, such as variations in the IR effectiveness measures, has a greater impact in terms of QPP outcomes than other factors, such as variations in the choice of IR models. An important outcome arising from this study is that further research on QPP should place greater emphasis on a clear specification of the experimental setup to enable better reproducibility.
In future we plan to expand our evaluations beyond the TREC Robust dataset. A natural question that we would like to explore concerns the impact of varying $\mathcal{Q}$ (the set of benchmark queries) on relative QPP outcomes.

\subsubsection*{\textbf{Acknowledgement.}}
The second and the fourth authors were supported by the Science Foundation Ireland (SFI) grant number SFI/12/RC/2289\_P2.

\bibliographystyle{splncs04}
\bibliography{main.bbl}

\end{document}